\documentclass[journal]{IEEEtran}
\renewcommand{\baselinestretch}{1.0}
\usepackage{times}
\usepackage{graphicx}
\usepackage[fleqn]{amsmath}
\usepackage{amssymb}
\usepackage{algorithm}
\usepackage[numbers]{natbib}
\usepackage{color}
\usepackage{multirow}
\usepackage{url}
\usepackage{array}
\newcommand{\PreserveBackslash}[1]{\let\temp=\\#1\let\\=\temp}
\newcolumntype{C}[1]{>{\PreserveBackslash\centering}p{#1}}
\newcolumntype{R}[1]{>{\PreserveBackslash\raggedleft}p{#1}}
\newcolumntype{L}[1]{>{\PreserveBackslash\raggedright}p{#1}}

\begin{document}

\title{Effective Super-Resolution Methods for Paired Electron Microscopic Images}

\author{Yanjun~Qian, Jiaxi Xu, Lawrence F. Drummy,
        and~Yu~Ding,~\IEEEmembership{Senior Member,~IEEE}
\thanks{Y. Qian is with the Department of Statistical Sciences and Operations Research, Virginia Commonwealth University, Richmond, Virginia 23220
 USA (e-mail: yqian3@vcu.edu).}
\thanks{J. Xu and Y. Ding are with the Department of Industrial \& Systems Engineering, Texas A\&M University, College Station, Texas 77843 USA (e-mail: xujiaxi@tamu.edu and yuding@tamu.edu).}
\thanks{L. Drummy is with the Materials and Manufacturing Directorate, Air Force Research Laboratory, Wright-Patterson Air Force Base, OH 45433 USA (e-mail: lawrence.drummy.1@us.af.mil).}
\thanks{This paper has supplementary downloadable material available at \protect\url{http://ieeexplore.ieee.org}, provided by the authors. The material includes the electron microscopic images and the MATLAB codes for the proposed method. Contact yqian3@vcu.edu for further questions about this work.}}

\markboth{Submit to {\it IEEE Transaction on Image Processing}}%
{Qian \MakeLowercase{\textit{et al.}}: Effective Super-Resolution Method for Paired EM Images}

\maketitle

\begin{abstract}
This paper is concerned with investigating super-resolution algorithms and solutions for handling electron microscopic images. We note two main aspects differentiating the problem discussed here from those considered in the literature. The first difference is that in the electron imaging setting. We have a pair of physical high-resolution and low-resolution images, rather than a physical image with its downsampled counterpart.  The high-resolution image covers about $25\%$ of the view field of the low-resolution image, and the objective is to enhance the area of the low-resolution image where there is no high-resolution counterpart.  The second difference is that the physics behind electron imaging is different from that of optical (visible light) photos.  The implication is that super-resolution models trained by optical photos are not effective when applied to electron images. Focusing on the unique properties, we devise a global and local registration method to match the high- and low-resolution image patches and explore training strategies for applying deep learning super-resolution methods to the paired electron images. We also present a simple, non-local-mean approach as an alternative.  This alternative performs as a close runner-up to the deep learning approaches, but it takes less time to train and entertains a simpler model structure.

\end{abstract}


%
\IEEEpeerreviewmaketitle

\section{Introduction}


In this paper, we consider an image processing problem encountered in nanomaterial characterization.  Material science researchers capture two-resolution electron microscopic (EM) images independently from the same specimen: a high-resolution (HR) image of $M\times N$ pixels, denoted by $\mathbf{I}_h$, and a low-resolution (LR) image, denoted by $\mathbf{I}_l$. The LR image has the same amount of pixels of the HR image but half of its resolution.

\begin{figure*}[!htbp]
\begin{center}
   \includegraphics [width=1.\textwidth]{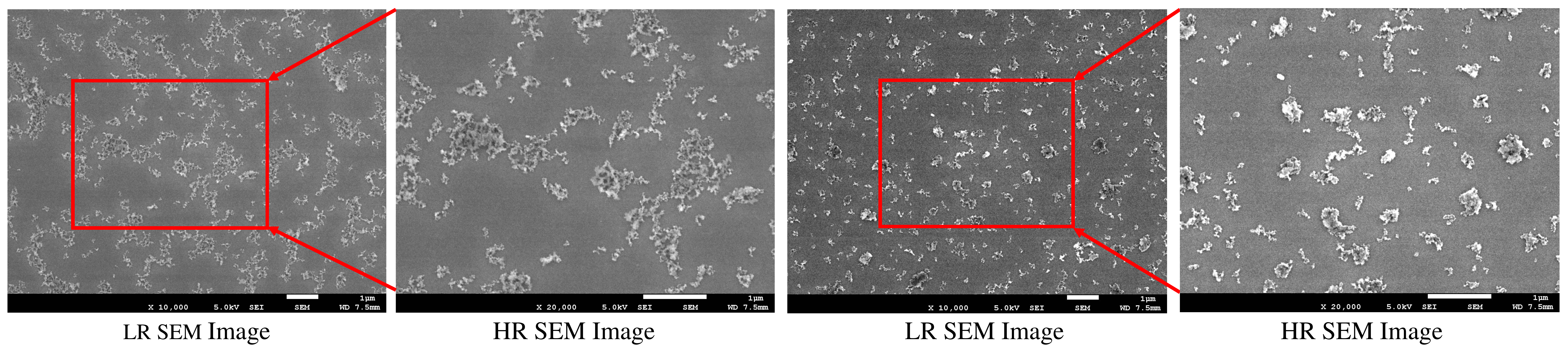}
\end{center}
   \caption{Two pairs of LR and HR SEM images. The red rectangles in the LR images are the areas corresponding to the HR images.}
   \label{fig:figure1}
\end{figure*}

Figure~\ref{fig:figure1} demonstrates two pairs of such EM images, both obtained by a scanning electron microscope (SEM). Each pair of images is obtained by the same SEM in one experimental setting but through two actions. First, the SEM is set at a low magnification level and takes the low-resolution image.  Then, with the same sample still in the specimen platform, the SEM is adjusted to a higher magnification level, i.e., it is zoomed in, and takes the high-resolution image. The view fields of the two images overlap completely, or more precisely, the high-resolution image covers a smaller subset of the view field of the low-resolution image.  The overlapping areas in the LR images in Figure~\ref{fig:figure1} are marked by the red rectangles. The objective is to develop a super-resolution (SR) method for reconstructing an HR image of $2M\times 2N$ pixels over the whole area that is covered by the LR image. The essence of the task is to enhance the area of the low-resolution image where there is no high-resolution counterpart. If a method can accomplish this research goal, material scientists can effectively survey a bigger area with imaging quality comparable to HR images but with less dense sampling.

HR images are desired for the purpose of material characterization because they capture and reveal fine structures of nanomaterials~\cite{park2012multistage, park2013segmentation, yang2014automatic, qian2016robust}.  But it is time consuming to capture HR images. While using a SEM or a transmission electron microscope (TEM), the images are created by an electron beam rastering through the material, so that the time cost will be at least proportional to the number of pixels. An equally important consideration is that the electron beam of an EM may damage the subtle structure of certain materials. Material scientists want to avoid dense sampling in electron imaging if at all possible. An effective SR approach, if available, can be of a great help to high-resolution electron imaging of materials.

In recent years, numerous SR methods have been proposed and reported \cite{park2003super, tian2011survey, yue2016image, timofte2017ntire}. We note two important differences, distinguishing the problem at hand from those considered in the literature. The first difference is that we have a pair of LR and HR images, both physical and obtained independently, rather than a physical HR image with its downsampled version. It is our understanding that most of the existing SR approaches in their default settings take the downsampled version of the HR images as the LR image inputs in their training. The second difference is that the physics behind electron imaging is different from that of optical photos taken under visible light.  The implication is that super-resolution methods trained by optical photos are not going to be effective when applied to electron images. We will provide quantitative evidence in Section \ref{sec:preliminary} to support our claim.


Focusing on these unique properties of our problem, we first examine how the existing methods perform while training with the physically captured image pairs. We test the two sets of popular SR methods: the sparse-coding based SR \cite{yang2010image, trinh2014novel} and deep-learning based SR \cite{kim2016accurate, lim2017enhanced, zhang2018image}. While the sparse-coding methods fail to yield satisfying results, we find that the deep-learning based approaches demonstrate a good degree of adaptability to our problem. Then, we propose a simpler SR method based on non-local means (NLM) \cite{buades2005non, sreehari2017multi}. A bit surprisingly, the NLM method performs rather competitively---as the closest runner-up and only slightly less effective than the deep learning-based SR.  The NLM method, on the other hand, is fast to train and has good interpretability, i.e., with a simple model straightforward to understand and a few tuning parameters. Having better interpretability allows clearer clues and easier adjustments for further improving a method (especially tailoring for specific applications) as well as double-checking to verify the soundness of certain outcomes for their consistency with domain science understanding and first principles.

We investigate different training strategies. We find that the self-training, in which the model is trained by the data from a specific pair of images, attains the most competitive performance for all methods, in spite of the limited number of training samples under such setting. This revelation appears to differ from the conventional wisdom in SR research, which prefers a large size of training samples even though some or many of the training samples are not directly related to the test image. Under self-training, we observe that simpler networks among the deep-learning approaches can produce SR results as competitive as complex networks but the training of the simple networks is much faster.  All these observations indicate that the strategy for super resolve the paired EM images is different from those for general SR problems.

The remaining parts of this paper are organized as follows. Section \ref{sec:preliminary} reviews the relevant literature and presents our preliminary analysis, which is to demonstrate that a model trained by an external, synthetic dataset is not effective for the paired image problem. In Section \ref{sec:framework} we first explain how to tailor the existing SR methods for the paired EM image problem. Then we present the simple NLM based SR method. Section \ref{sec:results} compares the performance of multiple methods and shows the benefits of the self-training strategy. In Section \ref{sec:conclusions}, we summarize our contributions and discuss possible extensions.

\section{Literature Review and Preliminary Analysis}
\label{sec:preliminary}

While the early SR literature focuses on restoring an HR image with multiple LR images (e.g., those in a short clip of video) \cite{park2003super, komatsu1993very, shah1999resolution}, the mainstream SR research nowadays is the single-image SR, starting with the seminal work by~\citet{freeman2002example} nearly twenty years ago. The idea of single-image SR is as follows. HR/LR patches are extracted from a set of training images, and a machine learning model is then built to map the images at the two resolutions. A test LR image will be segmented into overlapping patches, and the corresponding HR patches are to be inferred by the trained model. The HR image over the whole field of view is then reconstructed from these HR patches.

Numerous single-image SR methods for optical images have been proposed using different machine learning models. The neighborhood embedding (NE) algorithms \cite{chang2004super, chan2009neighbor, gao2012image} are based on the assumption that the HR and LR patches share similar manifold structures. An HR patch is estimated from the nearest neighbors of its LR counterpart in the manifold. The joint model methods \cite{sandeep2016single, huang2018single} learn a joint HR-LR patch distribution and predict HR images by maximizing the likelihood. The regression-based algorithms \cite{ni2007image, kim2010single, timofte2014a+, wang2016single} fit a regression model to map LR and HR patches and predict the HR patch using the LR patches as the regressors. The sparse-coding super-resolution (ScSR) methods \cite{yang2010image, trinh2014novel, yang2012coupled, wang2012semi} look for a parsimonious dictionary to encode the patches and reconstruct the HR patch from the coefficients of its LR counterpart. In recent years, the deep learning methods have been adopted for achieving single-image SR \cite{kim2016accurate, lim2017enhanced, zhang2018image, dong2014learning, zhang2018residual} and image restoration \cite{xu2014deep, schuler2015learning, xu2017motion}.  The deep-learning methods, e.g., very-deep super-resolution (VDSR) \cite{kim2016accurate}, enhanced deep-residual networks super-resolution (EDSR) \cite{lim2017enhanced} and residual channel attention networks (RCAN) \cite{zhang2018image}, achieve the best performance in recent single-image SR challenges \cite{timofte2017ntire, timofte2018ntire}.

In their default setting, ScSR and deep learning methods usually train their models from high-resolution optical images only.  An LR image is used but it is synthesized by blurring and downsampling the HR image. We refer to this type of LR images as the synthetic LR images. We design a preliminary experiment to demonstrate that this default setting is ineffective when applied to the paired EM images.

\begin{figure}[!htbp]
\begin{center}
   \includegraphics [width=1.\linewidth]{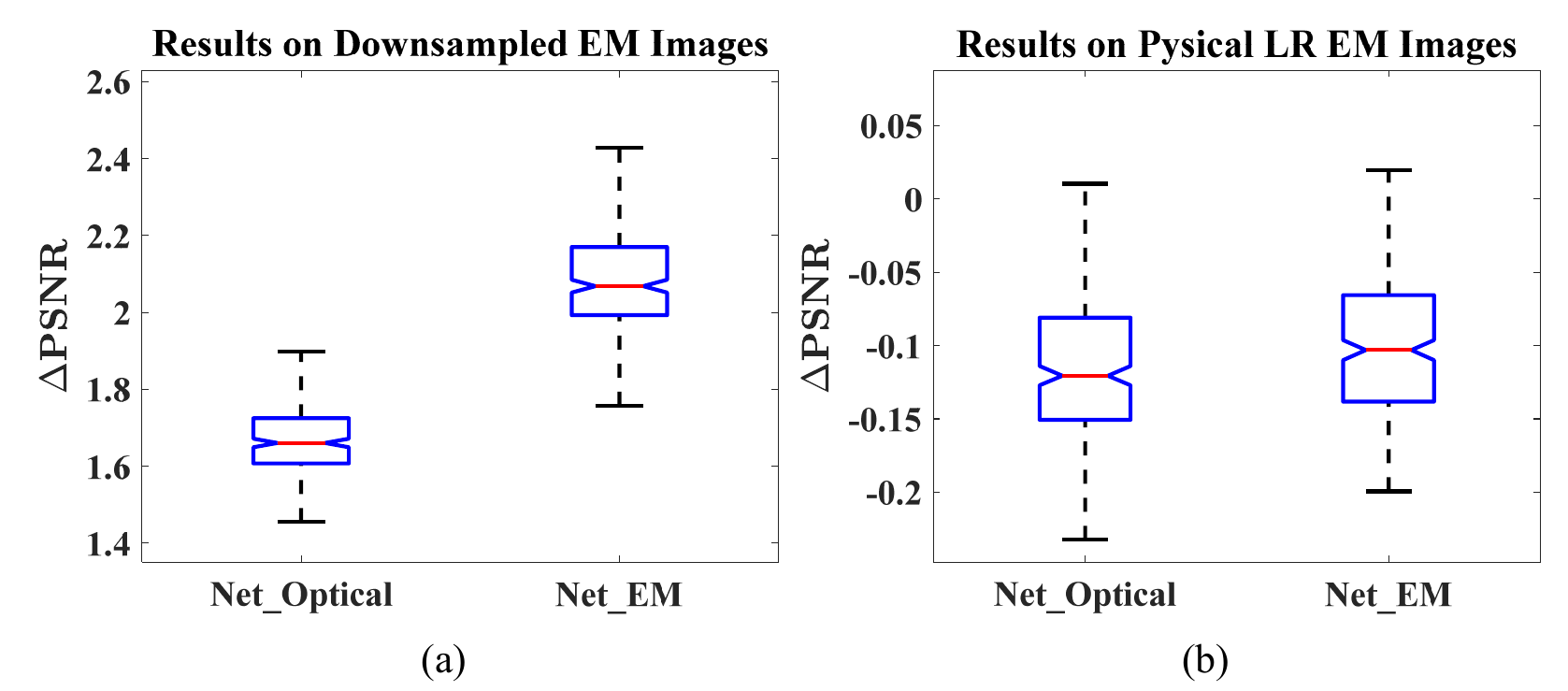}
\end{center}
   \caption{The performance of VDSR when its two versions, {\it Net\_Optical} and {\it Net\_EM}, are applied to (a) the downsampled EM images, (b) the physical LR EM images.}
   \label{fig:figure2}
\end{figure}

We train two networks with $41$ layers by VDSR using its default setting in \cite{kim2016accurate}: {\it Net\_Optical} from $539$ HR optical images from the IAPR TC-12 Benchmark \cite{grubinger2006iapr} , and {\it Net\_EM} from $539$ HR EM images collected by ourself. Then we test these networks using two datasets: synthetic electron images downsampled from the HR images and the physical LR electron images corresponding to the same HR images. After that, we compare the reconstructed images, presumably enhanced, with the actual HR images and calculate the peak signal-to-noise ratio (PSNR)---a high PSNR indicates a good reconstruction. Our baseline method is the bicubic interpolation \cite{keys1981cubic}, which is the most popular algorithm for upsampling an LR image to the pixel amount of the HR images. $\Delta$PSNR is computed as the difference between the PSNR of the image processed by a target method and the PSNR of the same image processed by bicubic interpolation.  $\Delta$PSNRs are shown in Figure~\ref{fig:figure2} when the two networks are applied to these two datasets.

One can make two observations from Figure~\ref{fig:figure2}. First, for both datasets, {\it Net\_EM} is more effective than {\it Net\_Optical}, highlighting the difference between optical and EM images. Second, although the two networks both work well on downsampled EM images (left panel), they perform worse relative to bicubic interpolation (right panel) when applied to the physical LR images. The message is that material scientists cannot simply grab an existing pre-trained SR model for processing the paired EM images. When we tried the sparse-coding SR methods \cite{yang2010image, trinh2014novel} or other deep-learning SR \cite{lim2017enhanced, zhang2018image} with their default setting, which use synthetic LR images, the resulting SR models are similarly ineffective.

\begin{figure*}[t]
\begin{center}
   \includegraphics [width=.85\textwidth]{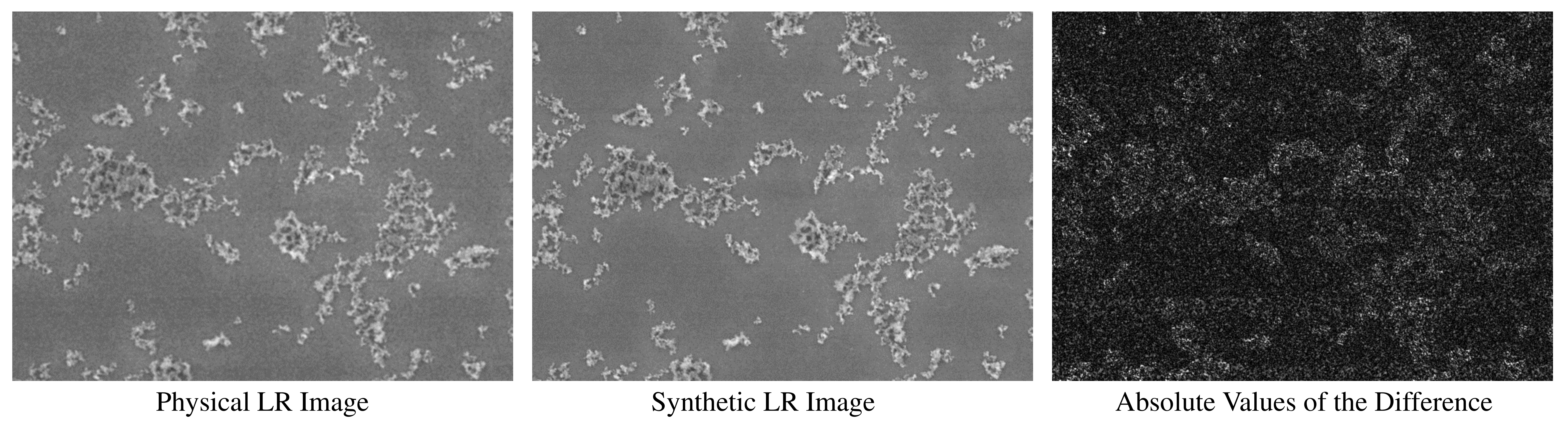}
\end{center}
   \caption{Comparison of a physical LR EM image and the synthetic downsampled image from the common corresponding HR image.}
   \label{fig:figure3}
\end{figure*}

In Figure \ref{fig:figure3}, we compare a physical LR image and a synthetic image, blurred and downsampled from their commonly paired HR image, and highlight their discrepancy. As we see in the right-most plot of Figure \ref{fig:figure3}, the difference between the two images is rather pronounced. We believe that the reason of discrepancy is in fact complicated, caused by the noise existing in the HR image, the different contrast levels between the paired images, and/or different natures and degrees of local distortion from individual image-capturing processes.

Although not studied in the SR literature as thoroughly as the single-image SR problems have been, there are in fact some initial attempts on the SR problem involving physically captured LR images. \citet{xu2017learning} develop a SR approach for blurred LR face and text images with a generative adversarial network (GAN). Then, \citet{xu2019towards} propose a framework to generate realistic training data from raw images captured by a camera sensor, and improve the SR results from the real LR optical images. \citet{zhang2019zoom} discuss super resolution trained with physical LR images. They improve the traditional deep learning based SR by using raw images from a camera and  introducing  a new loss function for handling the local distortions. Those works confirm and highlight the drawbacks of training from synthetic images when processing physical images, inspiring us to extend this line of research to paired EM images.

\citet{trinh2014novel} propose a SR method for paired medical images. To handle the noise in LR images, they store original image pairs to build a library and then solve a sparse representation for an input LR patch to reconstruct its HR counterpart. While accounting for noise in LR images, \citet{trinh2014novel} still ignore other discrepancies between the image pairs, e.g., the local distortion and differing contrasts. Their reconstruction process is also slow as solving the $L_1$ optimization for sparse representation is time-consuming.

\citet{sreehari2017multi} propose one of the first SR methods for EM images. In their approach, a library is built by an HR scan over a small field-of-view of a certain sample. When the LR EM image over a large field-of-view comes, a library-based non-local-mean method (LB-NLM) \cite{buades2005non} is applied to the upsampled LR image. After that, the HR image is recovered in a plug-and-play framework by invoking an alternating direction method of multipliers (ADMM) solver \cite{boyd2011distributed}. Compared with the SR methods for optical images, \citet{sreehari2017multi} build the library directly using electron image samples of nanomaterials, rather than unrelated optical images, and consider the noise in HR images explicitly in the plug-and-play framework. However, their algorithm does not include the physical LR images in the library, falling short of mapping the LR and HR patches directly. 



\section{Super-resolution Methods for Paired EM Images}
\label{sec:framework}

In this section, we proceed with two schools of approaches for handling paired EM images. The first school is to apply the current SR methods, specifically the sparse-coding methods \cite{yang2010image, trinh2014novel} and deep-learning methods \cite{kim2016accurate, lim2017enhanced, zhang2018image}, but using the physical EM image pairs as input. To handle the uniqueness of paired EM images, we explore different training strategies. The second school is to devise a simpler SR method that uses an LB-NLM filter with a paired library. The common preprocessing step in both schools is to register the HR and LR physical images; for that, we devise a global and local registration procedure.  The global registration is applied to the whole image, so that this step is common to all SR methods. The local registration is applied to the image patches and thus common only to the sparse coding methods and the LB-NLM method. The deep learning methods take the whole images as input to their networks and conduct an end-to-end super-resolution; for them, only is the global registration applied. In Section~\ref{sec:criterions}, we discuss the performance criteria used to evaluate the efficacy of the SR methods. Along with the commonly used PSNR and structural similarity (SSIM) \cite{wang2004image}, we also introduce some new metrics that we believe can articulate more pointedly the improvement made by the SR methods in the context of material characterization.


\subsection{Global and Local Registration}
\label{sec:registration}

With a pair of HR/LR EM images, $\mathbf{I}_h$ and $\mathbf{I}_l$, as inputs, we upsample $\mathbf{I}_l$ by a factor of two using bicubic interpolation; this produces $\mathbf{I}_u$, a $2M\times2N$ image. Then a shift transform $(x, y)$ and a rotation transform $(\theta)$ are applied to $\mathbf{I}_u$ and the mean squared error (MSE) between $\mathbf{I}_h$ and $\mathbf{I}_u$ are calculated in their overlapping area. We use a grid search to identify $(x, y, \theta)$ to globally minimize the MSE. To accelerate the searching process, we first downsample the two images by the same factor and roughly estimate $(x, y, \theta)$. Then we refine the estimation by searching its neighborhood using the original images. The registered upsampled image, denoted by $\mathbf{I}_r$, is transformed from $\mathbf{I}_u$ using the optimal global registration parameters.

\begin{figure}[!htbp]
\begin{center}
   \includegraphics [width=.8\linewidth]{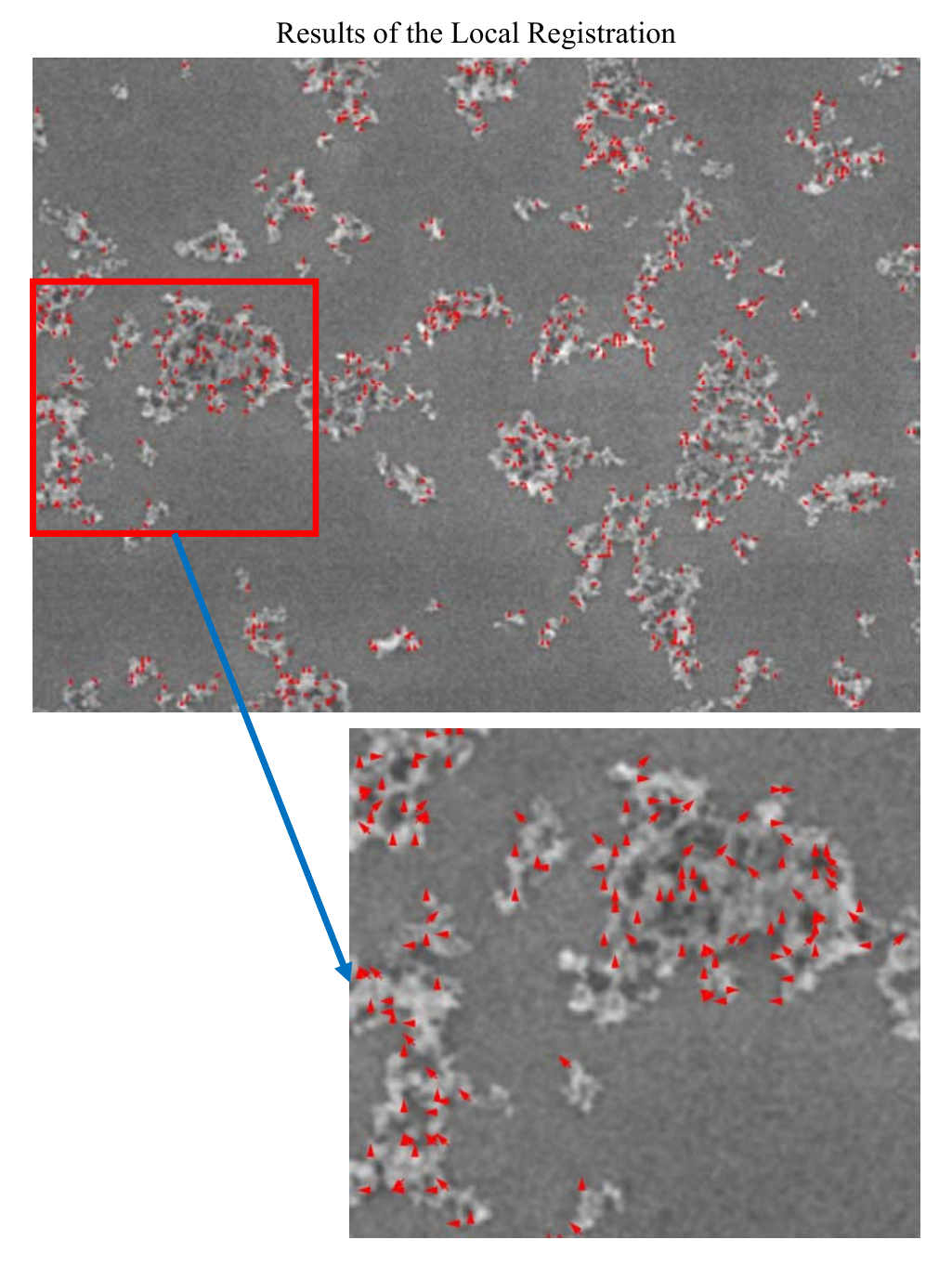}
\end{center}
   \caption{The results of the local registration. The bottom figure is magnified from the red rectangle in the top figure, in which the red arrows indicate the displacements $(i^*-i, j^*-j)$ between the matched patches.}
   \label{fig:figure4}
\end{figure}

To handle the local distortions between images, we segment the matched $\mathbf{I}_h$ and $\mathbf{I}_r$ into overlapping patches of size $n\times n$. $\mathbf{P}_h(i,j)$ and $\mathbf{P}_r(i,j)$ denote, respectively, the patches centered at $(i,j)$ in $\mathbf{I}_h$ and $\mathbf{I}_r$. Then we search the neighborhood of $(i, j)$ to find $(i^*, j^*)$ via solving the following optimization problem:
\begin{equation}
\min_{i^*,j^*}\frac{\mathbf{P}_h(i^*,j^*)\cdot\mathbf{P}_r(i,j)}{\Vert\mathbf{P}_h(i^*,j^*)\Vert_F\Vert\mathbf{P}_r(i,j)\Vert_F},
\end{equation}
where $\cdot$ denotes the inner product and $\Vert~\Vert_F$ is the Frobenius norm or the entrywise matrix 2-norm. We prefer the use of an inner product over the use of a Euclidean distance to match the two patches as the former is insensitive to the contrast difference between the two images. This criterion becomes less effective when the patches contain poor texture. Fortunately, the patches containing poor texture are the background patches, which are less important to the mission of super-resolution. We only apply the local registration to the patches with rich texture, which can be selected by deeming the variance of $\mathbf{P}_r(i,j)$ of a patch larger than a certain threshold. For our EM images, we set the threshold as $100$. Figure \ref{fig:figure4} presents one example after local registration, where the red arrows illustrate the displacements $(i^*-i, j^*-j)$ between the matched patches in $\mathbf{I}_h$ and $\mathbf{I}_r$. The magnitudes and directions of the displacements vary significantly across the image, showing a complex and irregular pattern of local distortions, which would not have been adjusted by a global registration alone.

\subsection{Existing SR Methods Applied to Paired Images}
After image registration, we can apply the popular SR methods to the paired EM images.  Here we test two main approaches: the sparse-coding methods and the deep learning methods.

When using the sparse-coding methods, we decide to remove the back-projection step after the SR reconstruction.  The back-projection step was included in the original sparse-coding method under the assumption that by downsampling the SR result, one can get the same image as the LR input. This assumption is not valid for the paired EM images; we articulated this point in Figure \ref{fig:figure3}. Our test shows that including the back-projection step deteriorates the SR result instead of improving it when the ScSR method is applied to the paired EM images.

When using the deep-learning methods, we are mindful of the small sample size of the paired EM images.  The small number of EM images is a result of the expensiveness to prepare material samples and operate electron microscopes. Acquiring \emph{paired} EM images would be even more time-consuming because doing so needs special care and specific experimental setup. In reality, one can expect a handful, to a few dozens at best, of such paired EM images. To prevent overfitting, we adopt two techniques: data-augmentation and early-stopping. A larger dataset is created by flipping each image pair row-wise and column-wise, rotating them by $90$, $180$ and $210$ degrees, and downsizing them by the factors of $0.7$ and $0.5$. By calculating the accuracy using validation data, we also discover that training achieves the best performance before its $30$th epochs and should be stopped accordingly.

There is the question of how to train a SR model.  The use of external image datasets for training, as done in the current SISR, is not the best practice in handling paired image problems, as shown in our preliminary analysis. Being ``external'', it means that the image pairs in the training set are unrelated to the image to be super resolved. That setting is understandable when one only has an LR image without its HR counterpart. For the paired image cases, given the complete overlap, albeit a subset of the view field, between an LR image and its HR counterpart, one would think that a relationship learned directly from this specific pair is the best for boosting the resolution for the rest of the LR image area uncovered by the HR image.

Suppose that we have a total of $m_\text{pr}$ pairs of SEM images, each of which has an LR image and its corresponding HR image. In this particular study, $m_\text{pr}=22$.  The size of both types of images is $1,280\times 944$ pixels. Through image registration, we identify the overlapping areas of each pair and carve out the corresponding LR image, which is of $640\times 472$ pixels. The $1,280\times 944$-pixel HR image and the $640\times 472$-pixel LR image are what we used to train the model and do the testing.  The non-overlapping area of the LR image is not used in the experimental analysis because there is no ground truth for that area to be tested.

\begin{figure*}[!htbp]
\begin{center}
   \includegraphics [width=.85\textwidth]{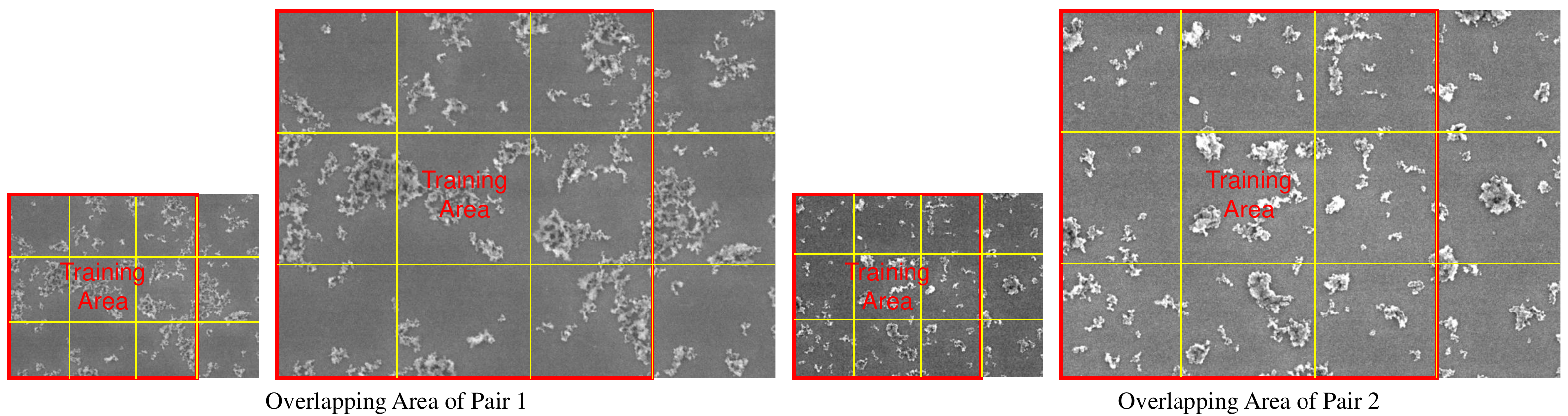}
\end{center}
   \caption{The overlapping areas of two pairs of SEM images. The left $75\%$ is the training area and the right $25\%$ is the test area. The yellow lines partition each image into $3\times 4$ subimages.}
   \label{fig:figure6}
\end{figure*}

To mimic the practical applications where the SR method is to be applied to the area where there is no corresponding HR images, we partition the LR and HR images in each pair into $3\times 4$ subimages.  We treat $m_\text{pp}^\text{tr}$ subimages as the training images and keep the remaining $m_\text{pp}^\text{ts}$ subimages unused in the training stage and treat them as the out-of-sample test images. In this study, the number of training images per pair is $m_\text{pp}^\text{tr}=9$ and the number of test images per pair is $m_\text{pp}^\text{ts}=3$.  The size of an HR subimage is $320\times 314$, where the size of an LR subimage is $160\times 157$, still maintaining the 2:1 resolution ratio. The training and test subimages of two SEM image pairs are shown in Figure \ref{fig:figure6}.

There are naturally two training strategies. To reconstruct the test subimages from Pair $i=1, \ldots, m_\text{pr}$, we can use the training subimages coming from the same pair to train the model. As such, there will be $m_\text{pr}$ individual models trained. In the phase of testing, each model is used individually for the specific image pair from which the model is trained. Each model is trained by $m_\text{pp}^\text{tr}$ pairs of subimages and evaluated on $m_\text{pp}^\text{ts}$ pairs of subimages.  We refer to this strategy as self-training.

Alternatively, we can pool all the training sample pairs together and train a single model. In the phase of testing, this single model is used for reconstructing the test images for all image pairs. We refer to this strategy as pooled-training. Under this setting, there are a total of $m_\text{pr}\times m_\text{pp}^\text{tr}$ pairs of training images and $m_\text{pr}\times m_\text{pp}^\text{ts}$ pairs of test images.  In the above example, the training sample size in the pooled-training is 198 pairs of subimages and the test sample size is 66 pairs of subimages, much greater than the sample sizes used in self training.

The conventional wisdom, especially when deep learning approaches are used, is that the $m_\text{pp}^\text{tr}$ training images, which are nine in this example, are too few to be effective. The popular strategy is to use the pooled training.  For the paired EM images, however, we find that using self-training in fact produces the best SR results, despite the relatively small sample size used. We believe this is something unique for the paired EM image problem---the pairing in the images makes using training samples internal to a specific image pair a better option than using more numerous external images.  We will present numerical evidences in Section~\ref{sec:results}.

\subsection{Paired LB-NLM SR Method for EM Images }
\label{sec:lb_nlm}

In this section, we propose a simple but effective SR method for the paired EM images, based on the LB-NLM filtering \cite{buades2005non, sreehari2017multi}. We build a paired library to connect the HR and LR patches from the training images. To include the informative patches for better training results, we design a clustering method for selecting the representative patches. The last step is a revised library-based non-local-mean (LB-NLM) method that reconstructs the HR images over the whole field of view, using the paired library of representative patches. The advantages of the LB-NLM method are its simple model structure and short training time, while its performance is less accurate only by a small margin than the deep learning based SR methods.

After the local registration, we store the matched patches from $\mathbf{P}_h$'s and $\mathbf{P}_r$'s into a paired library. \citet{sreehari2017multi} propose to create a library with dense sampling. The training area of each pair of EM images has about one million overlapping patches and many of them are of low texture and redundant information. We could, and should, reduce the library size to improve the learning efficiency.

As a large portion of the patches belongs to the background, random sampling is understandably not the most effective approach for patch selection. To ensure that different categories of image patches are adequately included, such as foreground, background, and boundaries, we devise a $k$-means clustering method to build the paired library, which is, in spirit, similar to the stratified random sampling approach as used in the design of experiments \cite{Wu2009}.

Assume that we would like to build a library with $L$ pairs of image patches, we randomly sample $K\times L$ HR patches from $\mathbf{P}_h$'s. Then we apply the $k$-means method to classify the HR patches into $k$ categories according to the vectorized intensity of the patches' pixels. After that, we randomly sample $L/k$ HR patches from each category, and store them and their matched patches $\mathbf{P}_r$'s in the library. We denote each pair of the patches by $\mathbf{P}_h^{(l)}$ and $\mathbf{P}_r^{(l)}$, respectively, for $l=1,\cdots,L$. When we choose a large enough $K$, say $10$, there are usually more than $L/k$ patches in each category. If the number of patches in one category is fewer than $L/k$, we can use all the patches in that category.  As a result, the library size is then smaller than $L$, but that is fine.

\begin{figure*}[!htbp]
\begin{center}
   \includegraphics [width=1.\textwidth]{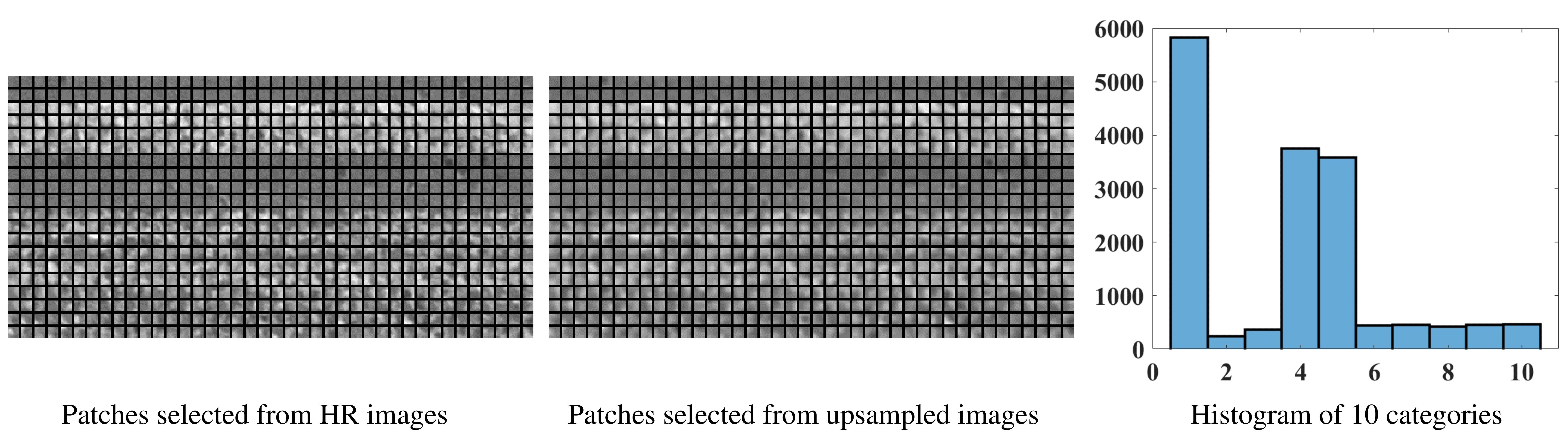}
\end{center}
   \caption{Demonstration of a paired library with $800$ patches of $9\times 9$, classified into $10$ categories. Left: the selected HR patches, where each row makes up one category; middle: the corresponding upsampled LR patches, right: the histogram of the original patches. }
   \label{fig:figure5}
\end{figure*}

In Figure \ref{fig:figure5}, we demonstrate a library with $800$ paired patches, each of size $9\times 9$. Figure \ref{fig:figure5}, the rightmost panel, presents the histogram of patches in $k=10$ categories of the original image data.  We can see that the first, fourth and fifth categories account for a large portion of the randomly sampled patches and these categories correspond to the patches in the background area. After the selection, there will be $80$ patches in each category equally. The background patches make up only $30\%$ ($3$ categories) of the selected ones in the library. The other $70\%$ ($7$ categories) are the patches with rich texture. Those $7$ categories include important diversity of the image information, which will play a critical role in the following SR step. Looking at the two figures on the left, one also observes that the noise and contrast levels are represented with a good balance in both HR and LR image patches.

With the paired library, we can reconstruct an HR image for the whole LR image area. In doing so, we first upsample, using bicubic interpolation and by a factor of two, the input LR image $\mathbf{I}_l$ to $\mathbf{I}_u$. Then we apply a revised LB-NLM filter to $\mathbf{I}_u$, based on the paired library established above, to obtain a filtered image $\mathbf{I}_f$. The filtered image $\mathbf{I}_f$ is the SR reconstruction of the physical LR image $\mathbf{I}_l$.

The revised LB-NLM filter runs as follows. For each pixel $(i,j)$ in $\mathbf{I}_u$, we extract an $n\times n$ patch $\mathbf{Q}_u(i,j)$ centered at $(i, j)$. Then a weight vector, $\mathbf{w}$, is calculated by comparing $\mathbf{Q}_u$ and the upsampled patches $\mathbf{P}_r^{(l)}$'s in the paired library as
\begin{equation}
w^{(l)}=\exp\left\{-\frac{||\mathbf{Q}_u(i,j)-\mathbf{P}_r^{(l)}||_2^2}{2n^2\sigma_n^2}\right\},
\label{eqn:weight}
\end{equation}
where $w^{(l)}$ is the $l$-th element of $\mathbf{w}$, while $\sigma_n$ controls the sparsity of the weight vector and can be interpreted as the assumed standard deviation of the image noise \cite{sreehari2017multi}. After $\mathbf{w}$ is normalized by $\mathbf{w}/\sum_{l=1}^Lw^{(l)}$, the reconstructed HR patch $\mathbf{Q}_h(i, j)$ is then calculated as the weighted average of the HR patches $\mathbf{P}_h^{(l)}$ in the paired library, such that
\begin{equation}
\mathbf{Q}_h(i, j)=\sum_{l=1} w^{(l)}\mathbf{P}_h^{(l)}.
\label{eqn:nlm}
\end{equation}
Then, the SR image $\mathbf{I}_f$ is reconstructed by combining $\mathbf{Q}_h(i, j)$ for all the $(i, j)$'s.

Since the patches in the library have been classified into $k$ categories, we accelerate the LB-NLM filter by calculating only the weights of the category closest to the current patch $\mathbf{Q}_u(i,j)$. As the weights are calculated by an exponential function, their values are close to zero when a category is dissimilar to the current patch. We compare the average value of the HR patches $\mathbf{P}_r^{(l)}$'s in each category with $\mathbf{Q}_u(i,j)$ to find the closest category. The selection is based on the shortest Euclidean distance between the average HR patch and $\mathbf{Q}_u(i,j)$.  Then the reconstructed HR patch $\mathbf{Q}_h(i, j)$ is obtained by the weighted average of the patches in this category alone. This approach can reduce the computational cost by $k$ times.

The scale parameter $\sigma_n$ can affect the LB-NLM filtering outcomes. For a small $\sigma_n$, $\mathbf{Q}_h(i, j)$ is determined by a few closest patches, yielding a similar result as the neighborhood embedding method \cite{chang2004super}. This line of method helps reconstruct image details in the foreground area. When $\sigma_n$ is large, LB-NLM averages a large number of patches in the library, decreasing the noise carried over from the training HR images. As EM images usually have a high noise level, especially in the background area, a default setting $\sigma_n=1.0$ can provide a good trade-off between enhancing signals and de-noising. 


We outline the steps of the paired library building and LB-NLM filtering in Algorithm \ref{tab:alg1}. By default, $n=9$, $k=50$, $K=10$, and $L$ is from a few thousand to tens of thousand.

\begin{algorithm}[!htbp]
\caption{The paired LB-NLM SR method. Inputs:  matched image patches $\mathbf{P}_h$'s and $\mathbf{P}_r$'s of size $n\times n$; LR image $\mathbf{I}_l$; parameters $k$, $K$, $L$, $\sigma_n$. Outputs: SR reconstruction, $\mathbf{I}_f$.}\label{tab:alg1}

{\bf Paired library building:}
\begin{enumerate}
\item Sample $K\times L$ patches from $\mathbf{P}_h$'s. Then use the $k$-means method to classify them into $k$ categories.
\item Sample $L/k$ patches from each category to obtain a library with $L$ HR patches.
\item Add the matched upsampled patch $\mathbf{P}_r$'s into the library. Each pair is denoted by $\mathbf{P}_h^{(l)}$ and $\mathbf{P}_r^{(l)}$.
\end{enumerate}
{\bf LB-NLM filtering:}
\begin{enumerate} \setcounter{enumi}{3}
\item Upsample $\mathbf{I}_l$ by a factor of two using bicubic interpolation.  The upsampled image is $\mathbf{I}_u$. Segment $\mathbf{I}_u$ into patches $\mathbf{Q}_u(i,j)$ of size $n\times n$.
\item Find the closest category to $\mathbf{Q}_u(i,j)$ in the library, which contains a subset of the patch indices, denoted by $C(i,j)$, in the library.
\item Calculate the weight $w^{(l)}$ for $l\in C(i,j)$:
\begin{equation*}
w^{(l)}=\exp\left\{-\frac{||\mathbf{Q}_u(i,j)-\mathbf{P}_r^{(l)}||_2^2}{2n^2\sigma_n^2}\right\}.
\end{equation*}
\item Normalize $\mathbf{w}$ by
\begin{equation*}
\mathbf{w}=\frac{\mathbf{w}}{\sum_{l\in C(i,j)} w^{(l)}}.
\end{equation*}
\item The reconstructed HR patch $\mathbf{Q}_h(i, j)$ is the weighted average of $\mathbf{P}_h^{(l)}$ as:
\begin{equation*}
\mathbf{Q}_h(i, j)=\sum_{l\in C(i,j)} w^{(l)}\mathbf{P}_h^{(l)},
\end{equation*}
\item Reconstruct the SR image $\mathbf{I}_f$ by combining $\mathbf{Q}_h(i, j)$ for all the positions $(i,j)$'s.
\end{enumerate}
\end{algorithm}

\subsection{Performance Criterions for Nanoimages}
\label{sec:criterions}

To measure the performance of an SR method, the most popular method is to consider the HR image as the ground truth, and compare it with the reconstructed image by calculating PSNR and SSIM. The closer the two images are, the higher PSNR and SSIM will be. Because bicubic interpolation serves as the baseline method, what is reported in the literature is $\Delta$PSNR or $\Delta$SSIM, i.e., the change made by a SR method over the bicubic interpolation baseline (as seen in Section \ref{sec:preliminary}).

As the foreground and background of EM images vary significantly, we also propose to segment the nanomaterial clusters (foreground) and the host material (background) through image binarization, and evaluate the improvements in PSNR and SSIM separately for the foreground as well as for the background. The foreground improvement reveals how well a SR method enhances the details of the image texture, whereas the background improvement points to a better de-noising capability.

The goal of super-resolution for EM images is to increase the ability of material characterization; for instance, increase the accuracy of morphology analysis. But PSNR and SSIM do not necessarily fully reflect a change in this capability.  Thus we propose to add a metric to measure more directly the impact made by an SR method, which is to check whether the reconstructed images are able to facilitate a better detection of nanomaterial's boundary. For that, we use Canny's edge detector \cite{canny1986computational} to identify the boundaries and textures of the nanomaterial clusters and label the detected edges in a binary map. Let $B_{\text{HR}}$ denote the binary edge map detected from the original HR image (ground truth) and $B_{\text{SR}}$ denote the binary map detected from the reconstructed image resulting from the proposed SR method. The similarity between them is defined as:
\begin{equation}
\text{sim}=1-\frac{\Vert B_{\text{HR}}\ne B_{\text{SR}}\Vert _1}{\Vert B_{\text{HR}}\Vert_1+\Vert B_{\text{SR}}\Vert_1},
\label{eqn:sim}
\end{equation}
where $B_{\text{HR}}\ne B_{\text{SR}}$ produces an indicator matrix whose element is $1$ where $B_{\text{HR}}$ and $B_{\text{SR}}$ have different values and $0$ otherwise, and $\Vert \cdot \Vert_1$ is the entry-wise matrix 1-norm. A high sim indicates a better performance.

\section{Experimental Results}
\label{sec:results}

\subsection{General Results of PSNR and SSIM}

With the two training options, self-training versus pooled-training, we test the following methods on the $22$ pairs of SEM images---ScSR \cite{yang2010image}, SRSW \cite{trinh2014novel}, VDSR \cite{kim2016accurate}, EDSR \cite{lim2017enhanced}, and RCAN \cite{zhang2018image}, the original LB-NLM method \cite{sreehari2017multi} and the paired LB-NLM method. The $22$ image pairs are partitioned into $198$ in-sample subimages and $66$ out-of-sample subimages. For ScSR, $L=80,000$ paired patches of size $9\times9$ are randomly sampled to train a paired dictionary of size $1,024$. The same paired patches also make up the library for SRSW. VDSR, EDSR and RCAN are trained with their default settings and the data-augmentation and early-stopping options. For the original and paired LB-NLM methods, a paired library with the same size as in SRSW is constructed using the corresponding portion of code in Algorithm \ref{tab:alg1}.

Table \ref{tab:tab1} presents the average improvement of PSNR and SSIM by these SR methods as compared with the bicubic interpolation baseline.  We also report the percentages of the failure cases, which are defined as when a SR result yields a negative $\Delta$PSNR.

The first observation is that for the paired image problem, self-training is a better strategy, despite the relatively small image sample size used. For all methods, self-training outperform pooled-training in terms of the out-of-sample $\Delta$PSNR. For most methods, the self-training also produces a better out-of-sample $\Delta$SSIM while for some methods the pooled-training's $\Delta$SSIM is better. But either way, the difference in $\Delta$SSIM is marginal. As we argue earlier, using the learned relationship specific to a particular image pair pays off when that relationship is used for reconstruction. This pair-specific information does not exist in the general single-image SR when an external training set is used.  Overall, self-training is indeed a better strategy because of its high accuracy and efficiency (training time is to be shown in Section \ref{sec:time}).

\begin{table*}[!htbp]
\centering
\caption{The improvements of PSNR and SSIM of the reconstructed SEM images after applying different SR methods, as compared with bicubic interpolation. The percentages of failure cases are also shown.}
   \label{tab:tab1}
\begin{tabular}{c|c|c|c|c|c}
\hline\hline
\multicolumn{2}{c|}{\multirow{2}{*}{}} & \multicolumn{2}{c|}{Self-Training} & \multicolumn{2}{c}{Pooled-Training}\\\cline{3-6}
\multicolumn{2}{c|}{} & In-Sample & Out-of-Sample & In-Sample & Out-of-Sample \\\hline
\multirow{2}{*}{ScSR \cite{yang2010image}} & $\Delta$PSNR & $0.26$ dB & $0.23$ dB & $0.18$ dB  & $0.19$ dB\\\cline{2-6}
& $\Delta$SSIM & $0.019$  & $0.015$  & $0.012$  & $0.014$ \\\cline{2-6}
& Failure cases & \multicolumn{2}{c|}{$7.2\%$} & \multicolumn{2}{c}{$1.2\%$}\\\hline
\multirow{2}{*}{SRSW \cite{trinh2014novel}} & $\Delta$PSNR & $1.41$ dB & $1.17$ dB &  $0.28$ dB  & $0.31$ dB\\\cline{2-6}
& $\Delta$SSIM & $0.026$ & $0.026$ & $0.019$ & $0.022$ \\\cline{2-6}
& Failure cases & \multicolumn{2}{c|}{$1.9\%$} & \multicolumn{2}{c}{$16.3\%$}\\\hline
\multirow{2}{*}{VDSR \cite{kim2016accurate}} & $\Delta$PSNR & $2.22$ dB & $2.07$ dB & $1.24$ dB  & $1.25$ dB\\\cline{2-6}
& $\Delta$SSIM & $0.052$  & $0.051$  & $0.044$  & $0.047$ \\\cline{2-6}
& Failure cases & \multicolumn{2}{c|}{$0\%$} & \multicolumn{2}{c}{$4.6\%$}\\\hline
\multirow{2}{*}{EDSR \cite{lim2017enhanced}} & $\Delta$PSNR & $2.16$ dB & $2.06$ dB & $1.56$ dB  & $1.35$ dB\\\cline{2-6}
& $\Delta$SSIM & $0.052$ & $0.052$ & $0.050$  & $0.051$ \\\cline{2-6}
& Failure cases & \multicolumn{2}{c|}{$0\%$} & \multicolumn{2}{c}{$4.5\%$}\\\hline
\multirow{2}{*}{RCAN \cite{zhang2018image}} & $\Delta$PSNR & $2.24$ dB & $2.07$ dB & $1.84$ dB  & $1.59$ dB\\\cline{2-6}
& $\Delta$SSIM & $0.053$ & $0.050$ & $0.051$  & $0.051$ \\\cline{2-6}
& Failure cases & \multicolumn{2}{c|}{$0\%$} & \multicolumn{2}{c}{$3.4\%$}\\\hline
\multirow{3}{*}{Original LB-NLM \cite{sreehari2017multi}} & $\Delta$PSNR & $0.46$ dB & $0.45$ dB & $0.23$ dB  & $0.28$ dB\\\cline{2-6}
& $\Delta$SSIM & $0.016$  & $0.016$  & $0.017$  & $0.018$ \\\cline{2-6}
& Failure cases & \multicolumn{2}{c|}{$4.2\%$} & \multicolumn{2}{c}{$10.6\%$}\\\hline
\multirow{3}{*}{Paired LB-NLM} & $\Delta$PSNR & $3.75$ dB & $1.67$ dB & $0.87$ dB  & $0.78$ dB\\\cline{2-6}
& $\Delta$SSIM & $0.132$  & $0.037$  & $0.034$  & $0.031$ \\\cline{2-6}
& Failure cases & \multicolumn{2}{c|}{$0\%$} & \multicolumn{2}{c}{$14.4\%$}\\
\hline\hline
\end{tabular}
\end{table*}


Among the methods in comparison, ScSR is not competitive when it is applied to the paired EM images. The lack of competitiveness of ScSR can be explained by certain options used in its training process. ScSR extracts the high-frequency features from LR images. As the physical LR images contain heavy noisy, those high-frequency features do not adequately represent the image information. Also, ScSR assumes the reconstructed HR patches sharing the same mean and variance as the input LR patches, which is again not true for the physically captured image pairs. SRSW, on the other hand, obtains much better results by directly using the original patches. However, the randomly sampled library used in SRSW retains too many background patches with very little useful information. Such construction of the image library hampers SRSW's effectiveness. This shortcoming is to be further highlighted in the foreground/background analysis in Section \ref{sec:other}.

Trained from the physically captured LR images, the performance of VDSR improves significantly as compared to the preliminary results in Section \ref{sec:preliminary}. In terms of both $\Delta$PSNR and $\Delta$SSIM, the three deep-learning methods yield very similar results under self-training. Using pooled-training, the most advanced RCAN achieves the best performance but still is beaten by its self-training counterpart. A possible reason is that RCAN can benefit in pooled-training from its complex architectures, but this advantage, however, disappears in self-training. Considering the training time cost (to be shown in Section \ref{sec:time}), VDSR under self-training appears to be the best candidate to the SR task for the paired EM images.


The simple, paired LB-NLM method achieves rather competitive performances and outperforms the original LB-NLM, ScSR and SRSW. There are certain similarities between the paired LB-NLM method and SRSW. The paired LB-NLM method accounts for more factors behind the difference between a pair of physical images acquired at different resolutions, whereas SRSW primarily deals with the noise aspect. Both LB-NLM and SRSW show an obvious better performance when applied to the in-sample images under self-training, while for ScSR, the deep-learning methods, and the original LB-NLM, the in-sample versus out-of-sample performance difference is much less pronounced.

The out-of-sample performance of the paired LB-NLM method under self-training reaches $80\%$ accuracy of the deep-learning methods under the same setting. Considering the simplicity of the paired LB-NLM method, it is difficult to imagine that a simple method like that is able to achieve such a performance, relative to deep learning methods, on the general single-image SR problems; these results highlight the uniqueness of the SR problem for paired EM images.

In terms of the failure cases, the paired LB-NLM method and the three deep-learning methods yield zero failure cases under self-training. Generally speaking, self-training produces fewer failure cases for all methods except ScSR than its pooled-training counterpart.



\begin{figure*}[!htbp]
\begin{center}
   \includegraphics [width=.9\textwidth]{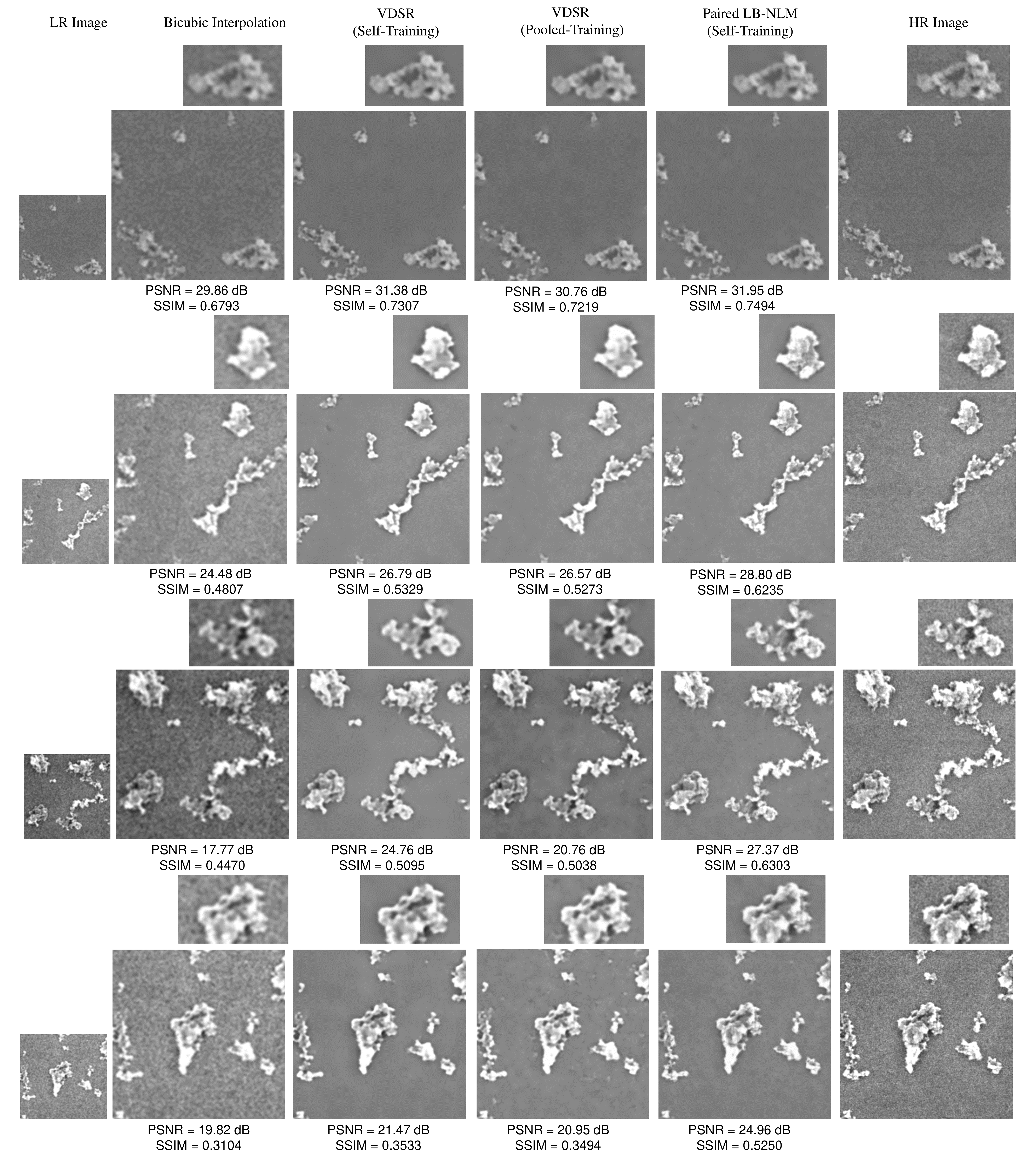}
\end{center}
   \caption{The LR images, the bicubic interpolated results, the image reconstruction results using VDSR (self-training and pooled-training), the paired LB-NLM method (self-training), and the ground truth (HR images) for four in-sample subimages. The small inserts in each row show a zoom-in of the foreground.}
   \label{fig:figure8}
\end{figure*}

\begin{figure*}[!htbp]
\begin{center}
   \includegraphics [width=.9\textwidth]{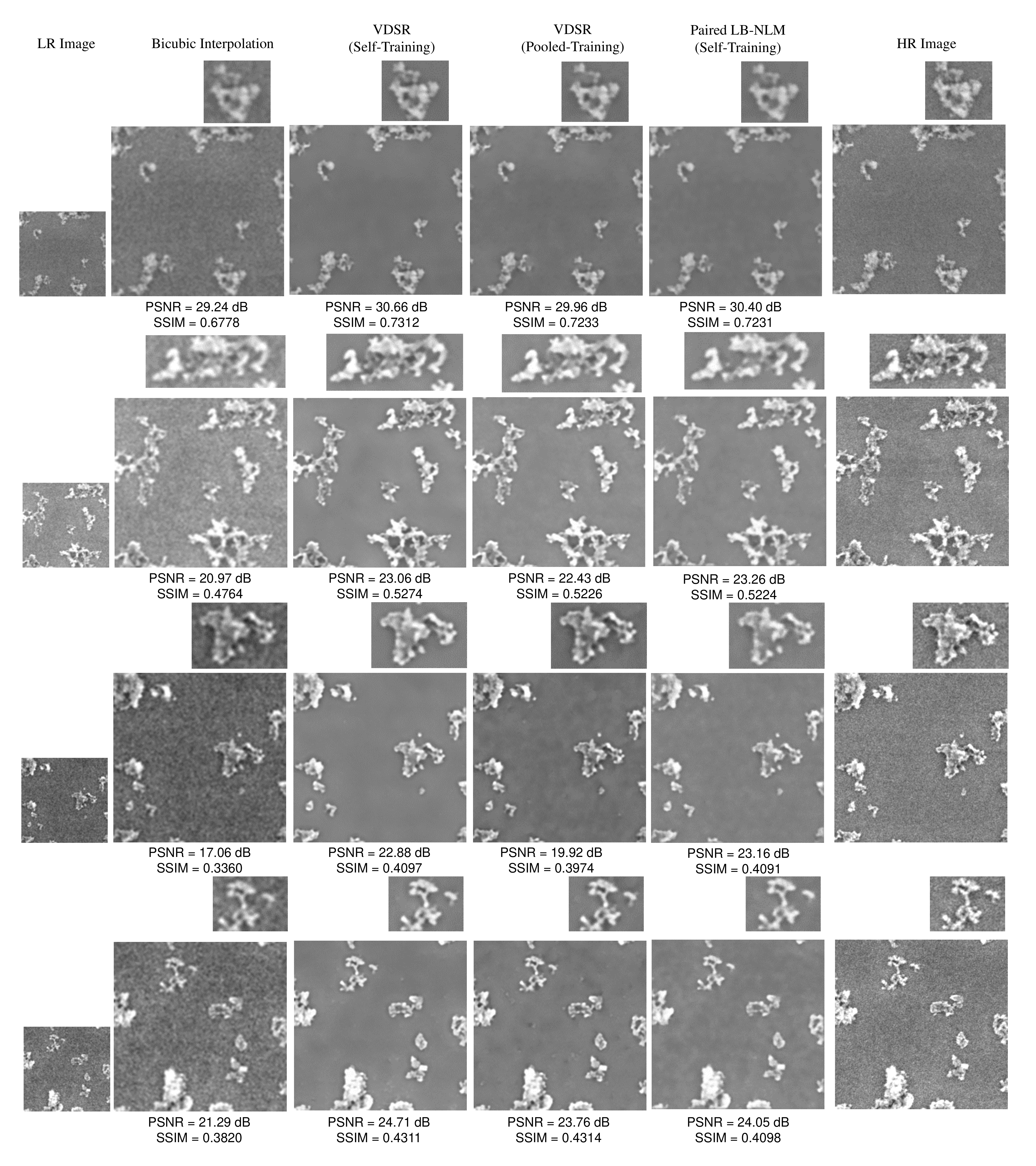}
\end{center}
   \caption{The LR images, the bicubic interpolated results, the image reconstruction results using VDSR (self-training and pooled-training), the paired LB-NLM method (self-training), and the ground truth (HR images) for four out-of-sample subimages. The small inserts in each row show a zoom-in of the foreground.}
   \label{fig:figure9}
\end{figure*}

We present in Figures \ref{fig:figure8} and \ref{fig:figure9} the original LR images, bicubic interpolated images, the reconstructed images by the VDSR (both self-training and pooled-training), the reconstructed images by the paired LB-NLM method (self-training only), and the HR images (ground truth). Here VDSR is used as a representative of the three deep-learning methods, since their respective best performances are similar. In each figure, four images are shown.  The four images in Figure~\ref{fig:figure8} are in-sample subimages, whereas those in Figure~\ref{fig:figure9} are out-of-sample subimages. VDSR and the paired LB-NLM method both give us a clear foreground and a less noisy background. The visual results of the LB-NLM method are comparable to those of VDSR under self-training. The visual comparison between the images under VDSR (self-training) and those under VDSR (pooled-training) highlights the benefit of using the self-training strategy---the benefit of using self-training is particularly noticeable on the last two images, namely the last two rows of Figures \ref{fig:figure8} and \ref{fig:figure9}.



Figures \ref{fig:figure8} and \ref{fig:figure9} also demonstrate the high noise level in both HR and LR EM images, which is a common phenomenon noted in the studies of EM images. The noise pattern is unique for each pair of EM images. For the paired image SR problems, the self-training strategy is more capable to learn the pattern and provide a more effective SR solution than pooled-training. It is worth noting that the background of the SR results from the three methods is clearer than that of the HR images. For the proposed LB-NLM method, it is a result of its inherent de-noising ability. As shown in Figure 6, the background patches in a library are close to each other. Thus, for an input patch from the background, the LB-NLM assigns similar weights $w^{(l)}$ to those library patches. The reconstructed results from equation (3) will be their average which has a lower noise level. VDSR shows similar results but the reason is not so easy to explain, due to the complexity involved in deep learning methods.

\subsection{Computation Time}
\label{sec:time}
We present the computation time of training and inference for five methods: three deep-learning methods, SRSW and the paired LB-NLM. We consider SRSW here as it is the better sparse-coding method. The three deep-learning methods, implemented by PyTorch, are trained at Texas A\&M University on one of its High Performance Research Computing (HPRC) Cluster with GPUs. The other two methods are trained with an MATLAB implementation on the same HPRC Cluster with parallel CPUs. Table \ref{tab:tab2} presents the average training and inference time when analyzing the 22 paired EM images.

With the aid of high computing power of GPUs, the deep learning methods still need a relatively long training time, especially when the pooled-training strategy is used. Training those models on a regular laptop computer without GPUs is not practical.  Both SRSW and the paired LB-NLM methods can be trained efficiently and used on regular personal computers.

Concerning the inference time in Table \ref{tab:tab2}, all deep learning methods run reasonably fast with GPUs, although RCAN is noticeably slower than the other two. We believe that the differences are caused by their network architectures. VDSR uses 20 layers and EDSR uses 69 layers, but RCAN uses more than $400$ layers. As RCAN adopts a much deeper network, its inference time becomes much longer. The paired LB-NLM's CPU time is similar to RCAN's GPU time. If all run on CPUs, LB-NLM is comparable or faster than the deep learning methods. SRSW suffers from a much longer inference time because it solves an $L_1$ optimization problem for each input LR patch.

As the self-training strategy produces the best results for the paired image SR problem, a user needs to re-train the model for every pair of newly captured images to attain the best enhancement. What this means is that unlike the traditional single-image SR problems, solving the paired SR problem prefers a method with a shorter training time. The long training time associated with the deep learning methods certainly puts them in a disadvantage. The proposed method, simpler and faster in training, presents itself as a competitive alternative, which can also be easily implemented on laptop computers without GPUs.




\begin{table*}[!htbp]
\centering
\caption{Computation time of training and inference of some SR methods on HPRC Clusters.}
   \label{tab:tab2}
\begin{tabular}{c|c|c|c|c|c}
\hline\hline
\multirow{2}{*}{} & \multicolumn{3}{c|}{PyTorch (training on GPU and inference on GPU or CPU)}  & \multicolumn{2}{c}{MATLAB (all on CPU)}\\\cline{2-6}
& VDSR \cite{kim2016accurate}  & EDSR \cite{lim2017enhanced} & RCAN \cite{zhang2018image} & SRSW \cite{trinh2014novel} & Paired LB-NLM \\\hline
\multirow{2}{*}{Training Time} & $\sim30$ mins (Self)  & $\sim30$ mins (Self)  & $\sim10$ hours (Self) & \multirow{2}{*}{$\sim5$ mins (Both)} &  \multirow{2}{*}{$\sim5$ mins (Both)}\\\cline{2-4}
& $\sim2$ hours (Pooled) & $\sim5$ hours (Pooled) & $\sim40$ hours (Pooled) & & \\\hline
\multirow{2}{*}{Inference Time} & $0.21$ sec on GPU & $0.17$ sec on GPU & $2.66$ secs on GPU & \multirow{2}{*}{$\sim 25$ mins}  & \multirow{2}{*}{$3.33$ secs} \\
& $4.08$ sec on CPU & $2.40$ sec on CPU & $21.65$ sec on CPU & &\\
\hline\hline
\end{tabular}
\end{table*}

\subsection{Further Performance Analysis}
\label{sec:other}

In this section, we provide quantitative analysis using the new criteria for EM nanoimages: the separate foreground/background analysis and the edge detection analysis.

\begin{figure*}[!htbp]
\begin{center}
   \includegraphics [width=.85\textwidth]{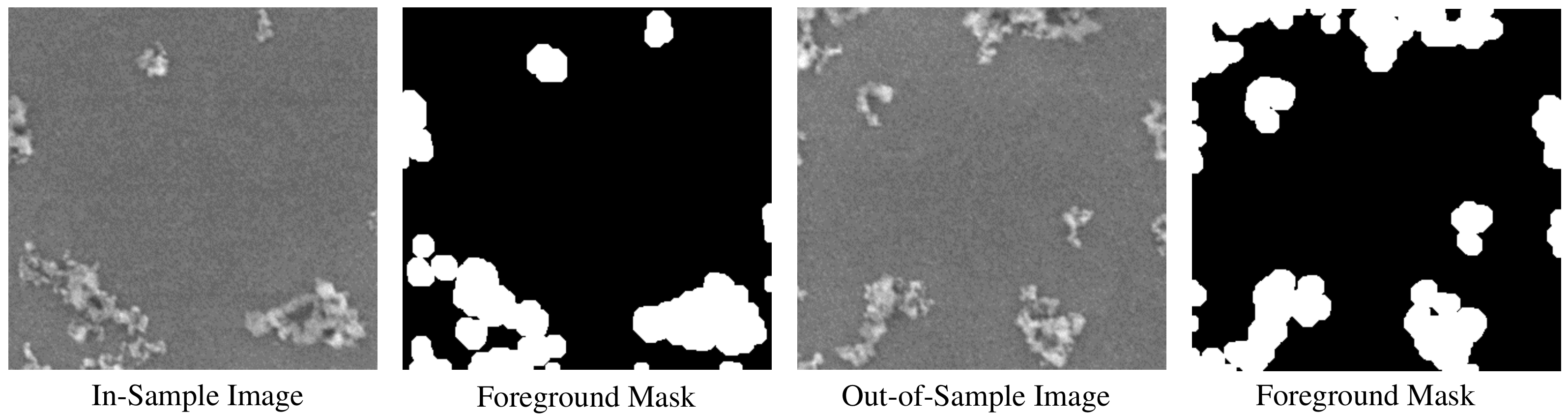}
\end{center}
   \caption{The foreground and background masks of an in-sample SEM subimage and an out-of-sample SEM subimage. The white areas indicate the nanomaterial (foreground), whereas the black areas indicate the host material (background).}
   \label{fig:figure10}
\end{figure*}

\begin{table*}[!htbp]
\centering
\caption{Changes in PSNR calculated for foreground and background for different SR results.}
   \label{tab:tab3}
\begin{tabular}{c|c|c|c|c|c}
\hline\hline
\multicolumn{2}{c|}{\multirow{2}{*}{}} & \multicolumn{2}{c|}{VDSR \cite{kim2016accurate}} & SRSW \cite{trinh2014novel} & Paired LB-NLM \\\cline{3-4}
\multicolumn{2}{c|}{} & Self-Training & Pooled-Training &  Self-Training & Self-Training \\\hline
\multirow{2}{*}{In-Sample} & Foreground & $1.21$ dB & $0.53$ dB & $0.03$ dB & $4.27$ dB\\\cline{2-6}
& Background & $2.86$ dB & $1.68$ dB & $2.27$ dB & $3.52$ dB\\\hline
\multirow{2}{*}{Out-of-Sample} & Foreground & $0.97$ dB & $0.48$ dB & $-0.25$ dB & $0.23$ dB\\\cline{2-6}
& Background & $2.83$ dB & $1.75$ dB & $2.15$ dB & $2.65$ dB \\
\hline\hline
\end{tabular}
\end{table*}

We first segment the SEM images by using Otsu's method \cite{otsu1979threshold} to highlight the separation of foreground from background and remove the isolated noise points in the foreground. Figure \ref{fig:figure10} shows the binary masks indicating the foreground versus the background in two images. Then we calculate separately the improvements of PSNR made by an SR method in the foreground and in the background.

Table \ref{tab:tab3} presents the changes in PSNR for three methods: VDSR (both self-training and pooled-training), SRSW (self-training), and the paired LB-NLM method (self-training). It is apparent that all these methods denoise the background much more than they enhance the foreground. The main advantage of VDSR is its ability to improve the foreground better than the paired LB-NLM.  This is not entirely surprising because the non-local-mean methods were originally designed as an image de-noising tool. It is also observed that the self-training VDSR is better than the pooled-training VDSR more so in terms of a stronger de-nosing capability over the background.  SRSW does a similar job in terms of denoising the background. But there is a slight decrease in terms of PSNR for the foreground, which suggests that the particular mechanism used in SRSW, especially the mechanism to create its library, is not effective for enhancing the foreground signals in the physical EM images.


\begin{figure*}[!htbp]
\begin{center}
   \includegraphics [width=.8\textwidth]{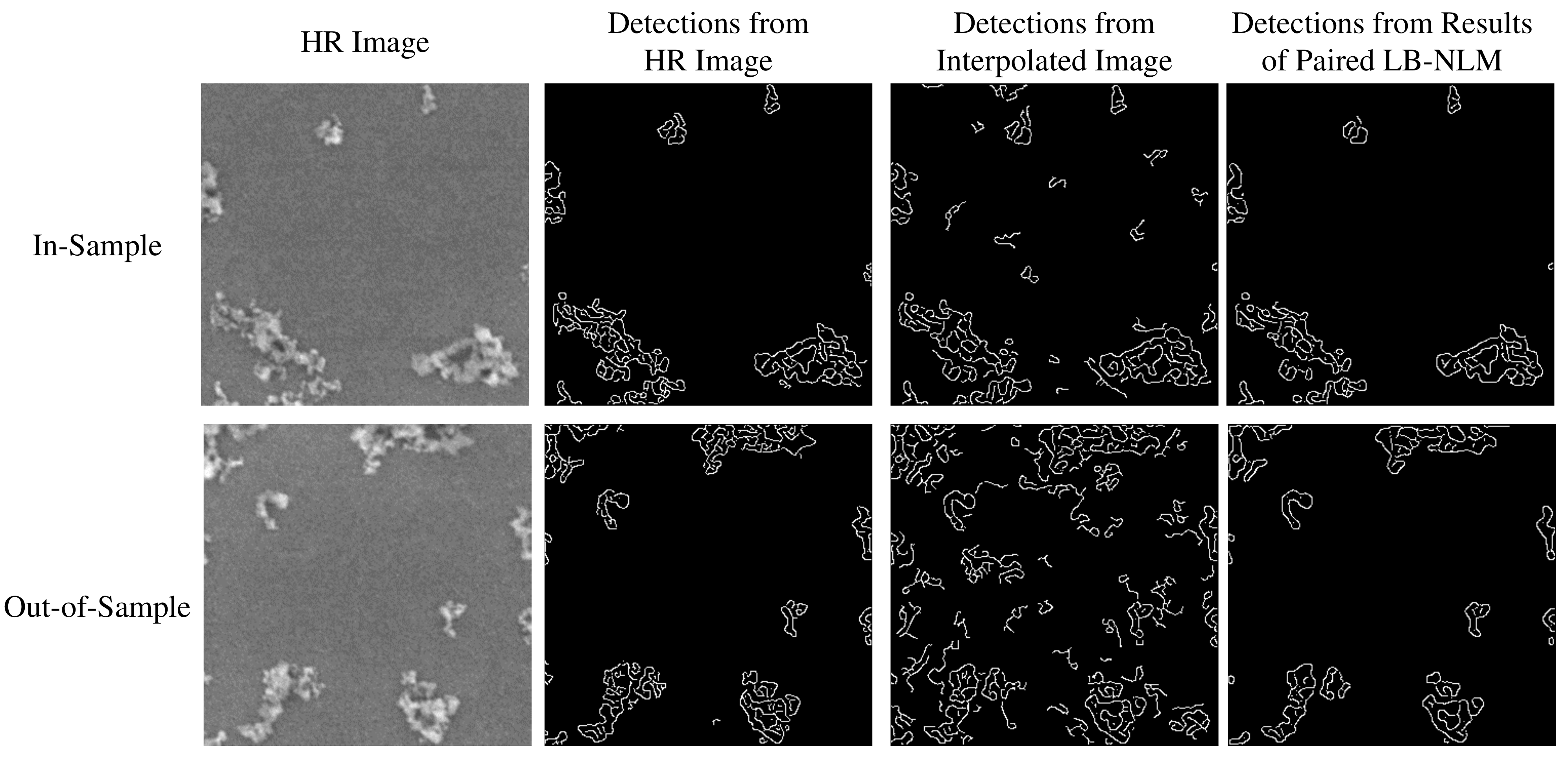}
\end{center}
   \caption{The results of Canny edge detection from the HR images and some reconstructed images.}
   \label{fig:figure11}
\end{figure*}

\begin{table*}[!htbp]
\centering
\caption{Results of sim for different SR methods and bicubic interpolation by Canny's detector.}
   \label{tab:tab4}
\begin{tabular}{c|c|c|c|c|c}
\hline\hline
\multirow{2}{*}{} & \multirow{2}{*}{Bicubic Interpolation} & \multicolumn{2}{c|}{VDSR \cite{kim2016accurate}} & SRSW \cite{trinh2014novel} & Paired LB-NLM \\\cline{3-4}
& & Self-Training & Pooled-Training &  Self-Training & Self-Training \\\hline
In-Sample & $0.25$ & $0.39$ & $0.37$ & $0.33$ & $0.56$ \\\hline
Out-of-Sample & $0.24$ & $0.37$ & $0.35$ & $0.24$ & $0.33$  \\
\hline\hline
\end{tabular}
\end{table*}

Next we apply Canny's edge detector \cite{canny1986computational} to the HR images, the bicubic interpolated images, and the reconstruction images by the three methods mentioned above. A key parameter in Canny's edge detector is set as $0.2$. Figure \ref{fig:figure11} demonstrates the detection results. The visual inspection show that the SR reconstructed results facilitate better edge detection than the bicubic interpolation baseline method.

To quantify the improvement in detection accuracy, we in Table \ref{tab:tab4} calculate the similarity index, sim, as defined in equation (\ref{eqn:sim}) in Section \ref{sec:criterions}. Except for SRSW, all methods can improve, as compared with the bicubic interpolation baseline, the Canny's detection accuracy by around $50\%$ on the out-of-sample images. The self-training VDSR achieves the largest improvement, although its sim is just slightly higher than that of the pooled-training VDSR and the paired LB-NLM. These results is consistent with the foreground PSNR improvements made by the four method in Table \ref{tab:tab3}.


\section{Conclusions and Future Work}
\label{sec:conclusions}
We present in this paper the paired EM image super-resolution problem and report our investigation of how best to address this problem.

Paired images are not very common in public image databases because taking them needs special care and specific setup. On the other hand, they are rather common in scientific experiments, especially in material and medical research.  The use of electron microscopes exacerbates the need for handling paired images for the purpose of super-resolution. Unlike optical photographing, the imaging process using an electron microscope is not non-destructive. Imaging using high-energy electron beams can damage sample specimen and must be carefully administrated. Consequently, researchers tend to use low-energy beams or subject the samples to a short duration of exposure. The results are of course low-resolution images. An effective super-resolution method that can subsequently boost these low-resolution images to a higher resolution has a significant impact on scientific fields relying on electron imaging.

In this research, we compare different state-of-the-art super-resolution approaches for handling the paired EM image problem. The take-home messages of our research can be summarized as follows:
\begin{itemize}
\item For the paired image problem, a local registration is important, as it accounts for the distortion between the image pairs. Our current approach is adequate but there is an ample room for further improvement.
\item When presented with paired images, it is recommended to use the self-training strategy, in spite of the relatively small sample size under that circumstance. Under self-training, simpler SR solutions are demonstrably as effective as more complex models.
\item  The paired LB-NLM entertains the advantage of fast training and simpler model structure and is a close runner-up to the deep-learning methods. It can be readily implemented on ordinary laptop computers.
\end{itemize}

Our work is among the very early efforts in addressing the paired EM image super-resolution problem. We see two broad future research directions as moving forward: (a) It is worthwhile to explore a deep neural network with a specialized architecture designed for the paired EM image problems. When the deep network can account for the two uniqueness in the problem (i.e., the image pairing and the electron images), a much greater enhancement of the low-resolution images can be anticipated; (b) It is interesting to observe the competitiveness of the simple, paired LB-NLM method or the simpler deep-learning network like VDSR. By exploiting the property and uniqueness of the paired image problems, it is possible to develop a computationally simple and structurally more interpretable method with good effectiveness.



\section*{Acknowledgment}

The authors would like to acknowledge the generous support from their sponsors. This work is partially supported by AFOSR DDDAS program grants FA9550-18-1-0144 and Texas A\&M X-grant program.

\ifCLASSOPTIONcaptionsoff
  \newpage
\fi

{\small
\bibliographystyle{IEEEtranN}
\bibliography{IEEEfull,QXDD_TIP_2019}
}

\begin{IEEEbiography}[{\includegraphics[width=1in,height=1.25in,clip,keepaspectratio]{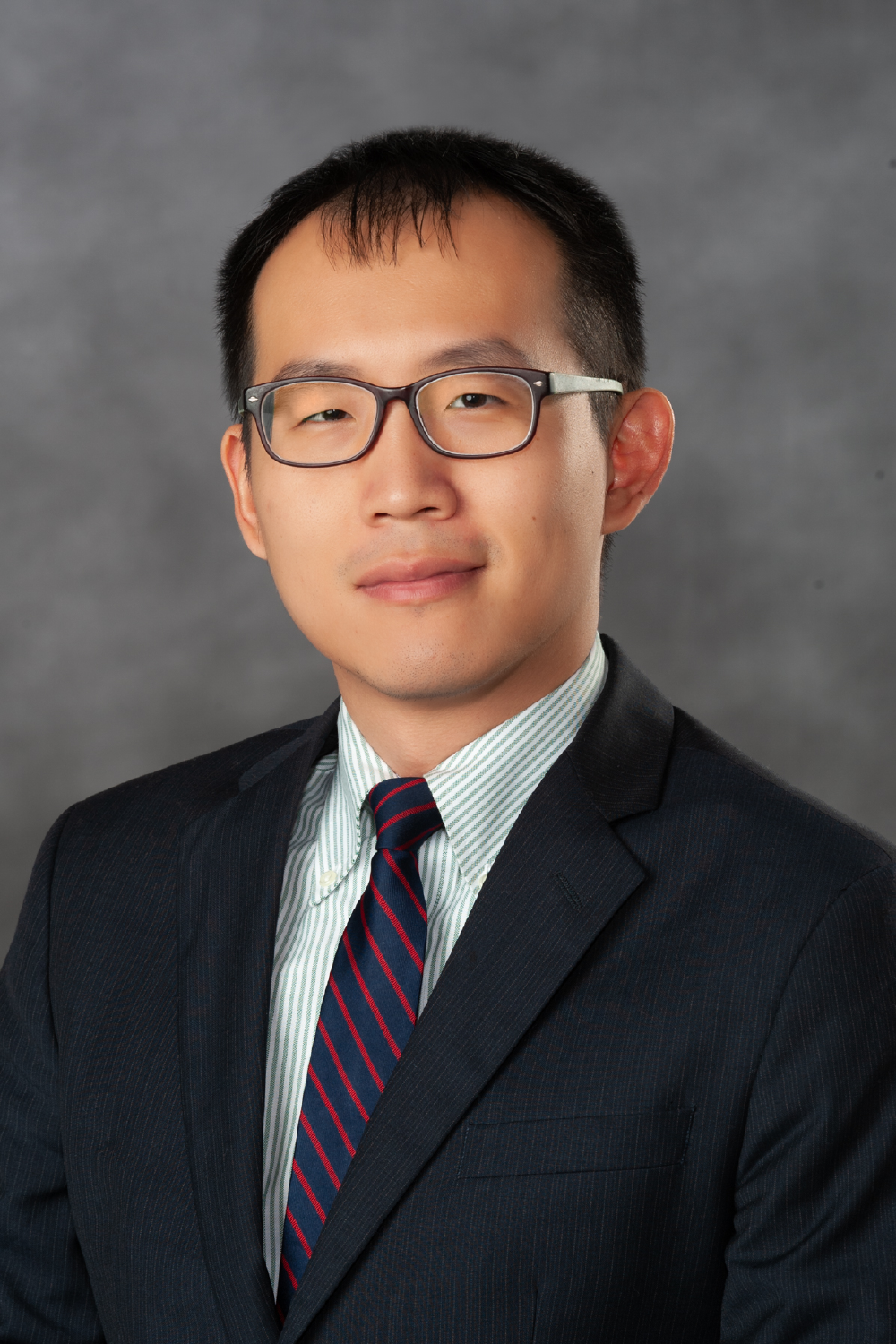}}]{Yanjun Qian} received B.S. (2009) and M.S. (2012) in Automation from Tsinghua University, China, and Ph.D. in Industrial \& Systems Engineering at Texas A\&M University (2018). He is currently an Assistant Professor of Statistical Sciences \& Operations Research at Virginia Commonwealth University. His research interests are in quality statistics and reliability, image and video processing, and machine learning. Yanjun Qian is a member of INFORMS, IISE, and SIAM.
\end{IEEEbiography}
\vspace{-1.5 in}
\begin{IEEEbiography}[{\includegraphics[width=1in,height=1.25in,clip,keepaspectratio]{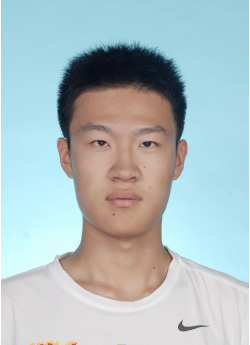}}]{Jiaxi Xu} received the B.S. in mechanical engineering from Shanghai Jiao Tong University, China, in 2017. He is currently a Graduate Research Assistant in Department of Industrial \& Systems Engineering, Texas A\&M University. His research interests include data science, machine learning and deep learning methodologies with emphasis on image processing applications.
\end{IEEEbiography}

\vspace{-1.5 in}
\begin{IEEEbiography}[{\includegraphics[width=1in,height=1.25in,clip,keepaspectratio]{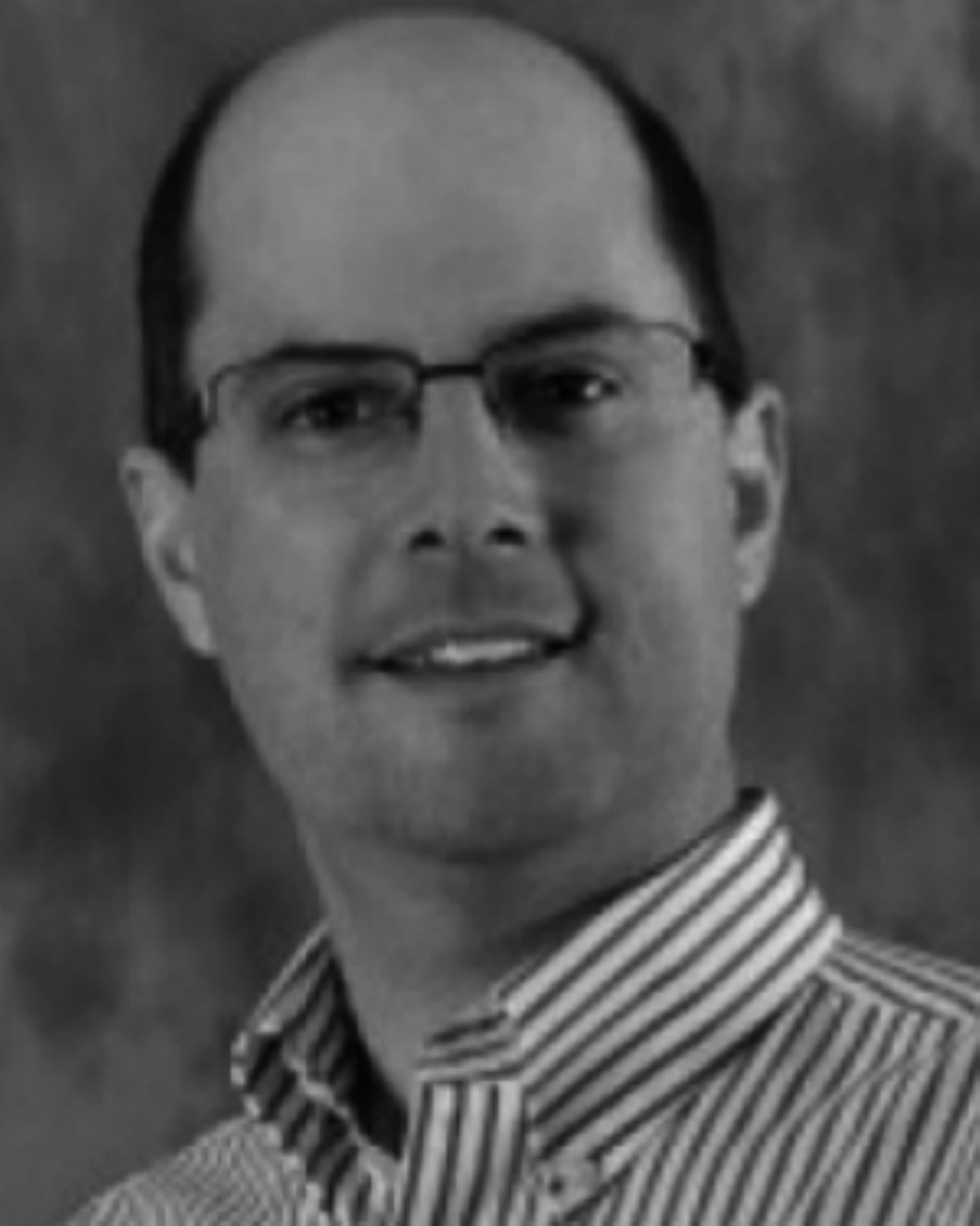}}]{Lawrence F. Drummy} received the B.S. degree in physics from Rensselaer Polytechnic Institute, Troy, NY, USA, in 1998, while developing image processing tools for scanning tunneling microscopy, and the Ph.D. degree from the Department of Materials Science and Engineering, University of Michigan, Ann Arbor, MI, USA, in 2003, while performing research on flexible electronic materials. He is a Materials Engineer with the Soft Matter Materials Branch, Functional Materials Division, Materials and Manufacturing Directorate, Air Force Research Laboratory, Dayton, OH, USA. His research interests include tomography and inverse problems, signal and image processing, high-resolution transmission electron microscopy, nanocomposites, organicinorganic interfaces, structural proteins, metamaterials, and organic electronics.
\end{IEEEbiography}

\vspace{-1.5 in}
\begin{IEEEbiography}[{\includegraphics[width=1in,height=1.25in,clip,keepaspectratio]{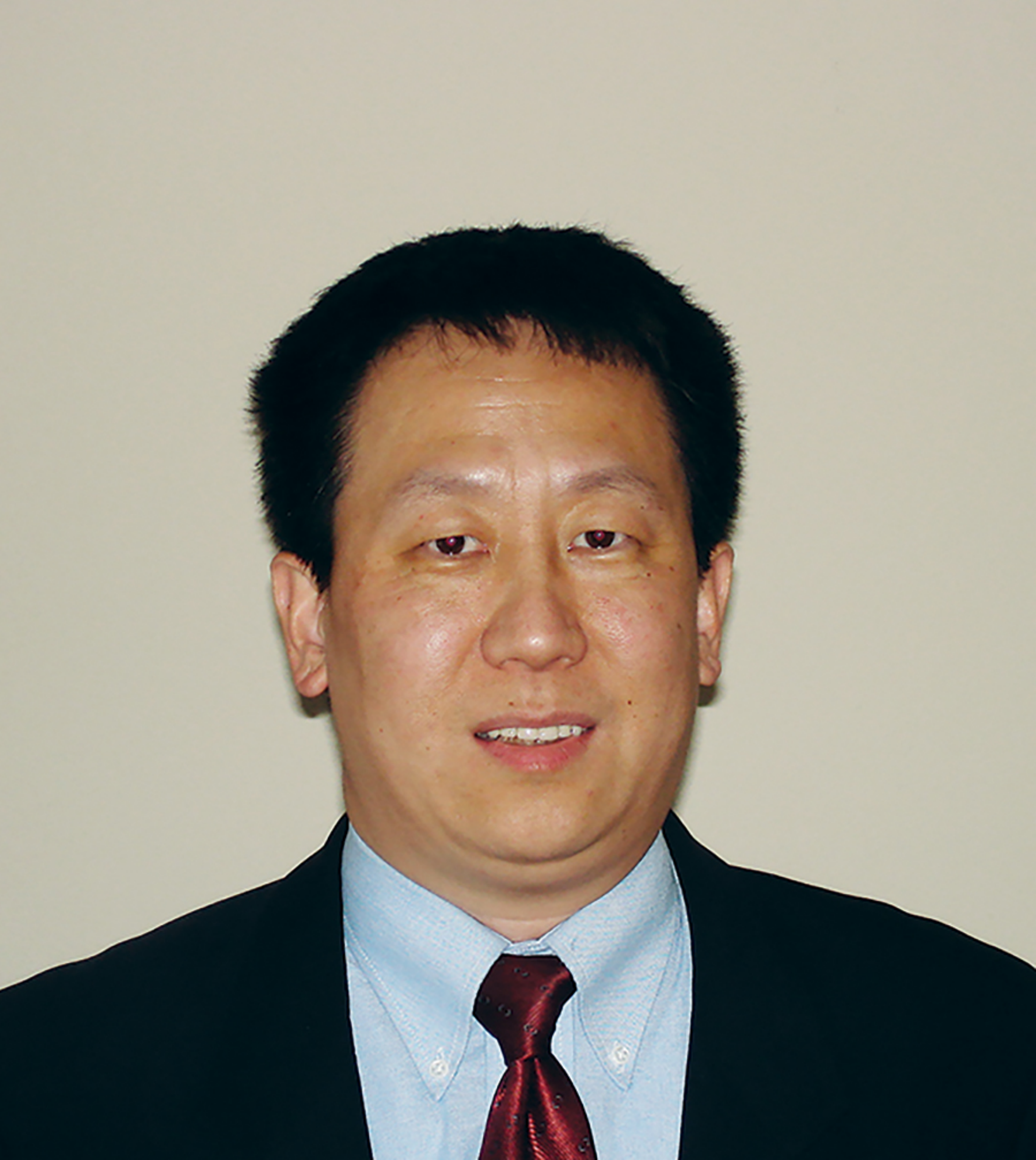}}]{Yu Ding} received B.S. from University of Science \& Technology of China (1993); M.S. from Tsinghua University, China (1996); M.S. from Penn State University (1998); received Ph.D. in Mechanical Engineering from University of Michigan (2001). He is currently the Mike and Sugar Barnes Professor of Industrial \& Systems Engineering, Professor of Electrical \& Computer Engineering at Texas A\&M University, and Associate Director for Research Engagement of Texas A\&M Institute of Data Science. His research interest is in the area of data and quality science. Dr. Ding is a recipient of the 2018 Texas A\&M Engineering Research Impact Award, the recipient of the 2019 Institute of Industrial and Systems Engineers (IISE)'s Technical Innovation Award, and a recipient of the 2020 AFS University-Level Distinguished Research Achievement Award in Research. He serves as an Editor for IEEE Transactions on Automation Science and Engineering, and is a fellow of IISE and ASME, a senior member of IEEE, and a member of INFORMS.
\end{IEEEbiography}

\end{document}